\documentstyle[aps,multicol,epsf]{revtex}
\begin{document}
\preprint{SNUTP }
\draft
\title{Anisotropic Surface Growth Model in Disordered Media \\}

\author{H.~Jeong$^{\dag}$, B.~Kahng$^{\ddag}$, and D.~Kim$^{\dag}$}
\address{
     \dag Center for Theoretical Physics and Department of Physics, \\
          Seoul National University, Seoul 151-742, Korea \\
     \ddag Department of Physics and Center for Advanced Materials and Devices,\\
           Kon-Kuk University, Seoul 143-701, Korea \\}

\maketitle
\thispagestyle{empty}

\begin{abstract}
We introduce a self-organized surface growth model in 2+1 dimensions 
with anisotropic avalanche process, which is expected to be 
in the universality class of the anisotropic quenched Kardar-Parisi-Zhang 
equation with alternative signs of the nonlinear KPZ terms. 
It turns out that the surface height correlation functions 
in each direction scales distinctively. The anisotropic behavior 
is attributed to the asymmetric behavior of the quenched KPZ 
equation in 1+1 dimensions with respect to the sign of the 
nonlinear KPZ term. \\  
\end{abstract}


\begin{multicols}{2}
\narrowtext

Subject of the pinning-depinning (PD) transition 
by external driving force has been of much interest recently. 
The problems of the interface growth in porous media under external 
pressure \cite{porous}, the dynamics of domain wall under random field
\cite{rfield}, the dynamics of charge density wave under external 
field \cite{cdw},  
and the vortex motion in superconductors under external current 
\cite{vortex,ertas} are typical examples. In the PD 
transition, there exists a critical value $F_c$ 
of the driving force $F$, such that
when $F < F_c$, interface (or charge, or vortex) is pinned 
by disorder, while for $F > F_c$, it moves with constant 
velocity $v$. The velocity $v$ plays the role of order 
parameter in the PD transition which behaves as   
\begin{equation} v \sim (F-F_c)^{\theta}.
\end{equation}

Recently, several stochastic models for the 
PD transition of interface growth in disordered media 
have been introduced~\cite{boston,tang}.
It is believed that the models in 1+1 dimensions, are described
by the quenched Kardar-Parisi-Zhang (QKPZ) equation
\begin{equation}
{\partial_t h }=\nu{\partial_x^2 h } 
+{\lambda \over 2}( \partial_x h)^2 +F+\eta(x,h),   
\end{equation}   
where the noise $\eta$ 
depends on position $x$ and height $h$ with the properties of
$\langle \eta(x,h) \rangle =0$ and 
$\langle \eta(x,h)\eta(x',h') \rangle =2D \delta(x-x')\delta(h-h')$.
The QKPZ equation exhibits the PD  
transition at $F_c$. The surface at $F_c$ 
can be described by the directed percolation (DP) cluster 
spanned perpendicularly to the surface growth direction 
in 1+1 dimensions. The roughness exponent $\alpha$ of 
the interface is given as the ratio of the correlation length exponents 
of the DP cluster in perpendicular and parallel directions, that is 
$\alpha=\nu_{\perp}/\nu_{\parallel}\approx 0.63$.\\  

The origin of the nonlinear term in the QKPZ equation 
is different from that of the thermal KPZ equation with the noise 
$\eta(x,t)$ \cite{dhar}. For the quenched case, the nonlinear term 
is induced by the anisotropic nature of disordered media, 
while for the thermal case, it is induced 
by lateral growth, and thus the coefficient $\lambda$ is proportional 
to the velocity of the interface, 
which vanishes at the threshold of the PD transition. 
For surfaces belonging to the DP universality class, 
the positive nonlinear KPZ term is induced under coarse graining 
of the quenched random force with amplitudes $\Delta_h^{1/2}$ and 
$\Delta_x^{1/2}$ in $h$-direction and in $x$-direction, respectively, 
when $\Delta_h > \Delta_x$.  
On the other hand, one may consider the case in 2+1 dimensions  
that the amplitudes of random force are anisotropic on substrate, that is,  
$\Delta_h > \Delta_{\parallel}$ in one direction of the substrate 
and $\Delta_h < \Delta_{\perp}$ in the other.
In such a case, following [8], the coarse-grained Langevin equation 
is expected to take the form of the anisotropic QKPZ (AQKPZ) equation 
given by
\begin{eqnarray}
\partial_t h=\nu_{\parallel}{{\partial}_{\parallel}}^2 h 
+ \nu_{\perp}{{\partial}_{\perp}}^2 h 
+{\lambda_{\parallel} \over 2}(\partial_{\parallel}h)^2
+{\lambda_{\perp} \over 2}(\partial_{\perp}h)^2 \nonumber \\
+F+\eta({\bf r},h), 
\end{eqnarray}  
with $\lambda_{\parallel} > 0$ and $\lambda_{\perp} < 0$. \\

In this letter, we study the surface of Eq.~(3) by introducing 
a self-organized stochastic model. The universality class of the 
stochastic model is checked by comparing surface properties with
those obtained from 
direct numerical integration of Eq.~(3). 
As shown in Fig.~1, the surface of Eq.~(3) at $F_c$ forms the shape of a mountain range
with steep inclination in one direction, whereas it is gently sloping 
in the other direction. Accordingly, the roughness exponents of 
the height-height correlation functions in each direction 
scale distinctively, which leads to a new universality class. 
This result is remarkable as compared with the case of the thermal noise. 
For the thermal case, the height-height correlation function is isotropic, 
and the anisotropic KPZ equation in 2+1 dimensions 
renormalizes into the Edwards-Wilkinson equation \cite{ew}.  
This is because for the case described by the KPZ equation with thermal noise,
the sign of the nonlinear terms is irrelevant in determining the universality
class \cite{kpz}, and 
the two nonlinear terms with different signs are canceled out 
effectively \cite{wolf}.  
However, for the quenched case, the anisotropic surface morphologies
of Fig.~1 imply that such cancellation does not occur and the QKPZ equation 
is asymmetric with respect to the sign of the nonlinear KPZ terms. 
Accordingly, in this letter, we also study the QKPZ equation 
with negative $\lambda$ in 1+1 dimensions, which enables one to 
understand the anisotropic nature of the AQKPZ equation 
in 2+1 dimensions.\\  
 
First, we study the QKPZ equation with $\lambda < 0$ in 
1+1 dimensions by direct numerical integration 
with the discretized version, 
\begin{eqnarray} 
h(x,&t&+\Delta t)=h(x,t)+\Delta t \{h(x-1,t)+h(x+1,t) \nonumber \\
&&-2h(x,t)+ {\lambda \over 8}(h(x+1,t)-h(x-1,t))^2+F \} \nonumber \\ 
&&+(\Delta t)^{2/3}\xi(x,[h(x,t)]),
\end{eqnarray} 
where [$\cdots$] denotes the integer part, and $\xi$ is
uniformly distributed in [${-{1 \over 2}}$, $1 \over 2$].
The prefactor $(\Delta t)^{2/3}$ of the noise term arises from approximately
coarse-graining the noise $\eta(x,h)$ during the time interval $\Delta t$.
Numerical integration using Eq.~(4) for $\lambda=1$ and $\Delta t=0.01$ yields,
even for a modest system size of $L=10^3$, the roughness exponent $\alpha \approx 0.63$,
which is consistent with the value of the DP universality. 
Note that the use of usual prefactor $(\Delta t)$ requires 
much larger computational cost to obtain the
DP value of $\alpha$ \cite{vicsek,leschhorn}.
Details for derivations of Eq.~(4) 
and result of the direct numerical integration for $\lambda>0$ will be 
published elsewhere \cite{noise}. \\

For the case of $\lambda < 0$, we performed 
the numerical integration with $\lambda=-1$ and $\Delta t=0.01$ 
for convenience. Fig.~2 shows typical surface 
configurations evolved temporally at the PD  
transition point, $F_c \approx 1.98$, which exhibits the shape of 
a mountain with flat inclination in the pinned 
state. The surface in the pinned state looks similar to that 
of the model A by Sneppen \cite{sneppen}, and 
the shape of the surface determines the roughness exponent 
to be $\alpha=1$.  
Note that the RSOS restriction before deposition in the model A 
of Sneppen does not allow the particle to deposit in every site, 
which makes the growth velocity reduced, and results in $\lambda <0$ 
\cite{park}. Accordingly, it is reasonable to have the surface 
morphology as shown in Fig.~2 for the QKPZ equation 
with negative $\lambda$, which is different from the 
one for $\lambda > 0$. In addition, we also 
examined the noise distribution on perimeter sites.  
It reveals that the pinning is caused by relatively 
large pinning strengths around the site where the height is minimum. 
The flatness on inclination makes the term $(\partial_x h)^2$ large, 
which with the large pinning strengths compensates the external 
driving force. Thus the growth velocity becomes zero, and the surface 
is pinned.  Since the surface pinning is caused mainly by the barrier 
of the pinning strengths around the site of minimum height, 
the growth velocity exhibits a sudden jump as the barrier 
is overcome by increasing the external driving 
force $F$. Accordingly, the PD transition is of first-order as 
depicted in Fig.~3. On the other hand, for $\lambda > 0$, 
pinned sites are not localized, but scattered, 
so that the PD transition is continuous.
Therefore the surface of the QKPZ equation in 1+1 dimensions 
is asymmetric with respect to the sign of the nonlinear term. \\  

Next, we study the AQKPZ equation, Eq.~(3), 
in 2+1 dimensions by introducing 
a stochastic model.  It is a natural extension 
of the RSOS model introduced previously by the current authors 
to study the anisotropic thermal KPZ equation \cite{jeong}. 
The stochastic model is based on a combination of the Sneppen's model
A for $\lambda <0$ and the model B for $\lambda > 0$ 
which is realized by assigning an anisotropic avalanche process. 
The advantage of studying such stochastic model is twofold; first there
is no need for fine-tuning of $F$ to get the critical state and second, asymptotic
states can be readily reached for small system size.
The model is defined on the checkerboard lattice, 
square lattice rotated by 45$^\circ$. 
Initially we begin with a flat surface characterized by the 
heights 0 on one sublattice and 1 on the other (see Fig.~4). 
Random numbers are assigned to each site.  
At each time step, selected is the site with minimum random 
number among the sites $(i,j)$ of which two nearest neighbors 
at $(i+\frac{1}{2},j-\frac{1}{2})$ and $(i-\frac{1}{2},j-\frac{1}{2})$ 
are higher. The site is updated by increasing the height by 2.
Next, the anisotropic avalanche process may occur on the neighboring 
sites, $(i+\frac{1}{2},j+\frac{1}{2})$ and 
$(i-\frac{1}{2},j+\frac{1}{2})$.  
If their height is lower by 3 than 
that of $(i,j)$, then the height is increased by 2. 
The avalanche rule is then applied successively to next rows 
in $\hat{j}$-direction until there is no change. The sites with 
increased height are updated by new random numbers.
The avalanche direction, $\hat{j}$-direction, corresponds to
the $r_{\parallel}$-direction in Eq.~(3).
The anisotropic avalanche process along positive $\hat{j}$-direction
is an interesting aspect of our model, which is 
a generalization of the model by Maslov and Zhang 
\cite{maslov} into 2+1 dimensions. Such anisotropic avalanche process
is to be distinguished from the isotropic avalanche process on tilted
substrates, in which the roughness exponent along the tilt direction 
($r_{\perp}$-direction) is 1/3 \cite{dhar}. 
Using the tilt argument, it can be shown 
that our model includes alternative signs of the nonlinear 
terms, that is, 
$\lambda_{\parallel} > 0$ and $\lambda_{\perp} < 0$ \cite{jeong}. 
A typical surface morphology is shown in Fig.~1a. 
We measured the roughness exponents for the height-height 
correlation functions, $C_{\parallel} ({r_{\parallel}}) 
\equiv \langle {1 \over L^2} \sum_x (h({\bf{x}})-h({\bf{x}}+
{\bf{r_{\parallel}}}))^2 \rangle 
\sim r_{\parallel}^{2\alpha_{\parallel}}$, and 
$C_{\perp}(r_{\perp}) \sim r_{\perp}^{2\alpha_{\perp}}$.  
The roughness exponents for each direction are obtained as 
$\alpha_{\parallel}=0.25(1)$ and $\alpha_{\perp}=0.75(1)$ 
as shown in Fig.~5. 
We also measured the height fluctuation width, 
$W^2 \equiv {1 \over L^2}\sum_r (h_r - \bar h)^2 
\sim L^{2\alpha}$ for $t \gg L^z$ and $\sim t^{2\beta}$ 
for $t \ll L^z$, where the exponents $\alpha$, $\beta$, and 
$z$ are the roughness, the growth, and the 
dynamic exponents, respectively, and $L$ is system size. 
It is obtained that $\alpha=0.87(1)$, $\beta=0.80(1)$, 
and $z$ is given as $z=\alpha/\beta$.
The values of the exponents are different from those of the 
isotropic case in 2+1 dimensions 
where $\alpha \approx 0.48$ and 
$\beta \approx 0.41$ \cite{amaral}.  
We also measured the avalanche size distribution, 
$P(s)\sim s^{-\tau}$. The exponent $\tau$ is obtained 
as $\approx 1.35(3)$.\\ 

In order to check if the stochastic model reduces to the 
AQKPZ universality, we considered the surface of Eq.~(3) 
by carrying out the direct numerical integration using the two dimensional
version of Eq.~(4). 
We used the numerical values of $\Delta t = 0.01$, $\lambda_{\parallel}=1$, 
and $\lambda_{\perp}=-1$ for convenience. 
Fig.~1b shows the surface morphology obtained by the 
direct numerical integration at the threshold of the 
PD transition, $F_c \approx 0.50$, which looks similar to the one 
in Fig.~1a. 
However, we could not measure the roughness exponents 
$\alpha_{\parallel}$ and $\alpha_{\perp}$ precisely, 
because their precise measurement requires relatively 
large system size and huge computing times.  
Nevertheless, since the morphologies of Fig.~1a and 1b are similar to 
each other and that of $\lambda_{\parallel} > 0$ and $\lambda_{\perp} < 0$ can be proven using 
the tilt argument for the stochastic model \cite{jeong},   
we believe that the stochastic model belongs to the AQKPZ universality.
The PD transition turns out to be continuous
as depicted in Fig.~6. The velocity exponent $\theta$ 
defined in Eq.~(1) is obtained as $\theta=0.9(1)$ 
which is somewhat larger than $\theta \approx 0.8$ 
for the isotropic case \cite{amaral}. 
The numerical results, $\tau \approx 1.35$, $z \approx 1.09$, and 
$\theta \approx 0.9$ seem to represent the characteristics of the 
self-organized critical depinning transition \cite{maslov,pmb}, 
but the relation of those exponents to the anisotropic 
roughness exponents in 2+1 dimensions is not clear yet. 
Further study is required about this point. 
We have also examined the surface in moving state, $F > F_c$. 
In this regime, the surface is no longer anisotropic, 
and reduces to the AKPZ equation with thermal noise for $F \gg F_c$. \\ 
  
In summary, we have studied the AQKPZ equation 
with alternative signs of the nonlinear terms in 2+1 dimensions, 
and found that it leads to a new universality class. 
The surface exhibits anisotropic scaling behavior, which is 
due to the asymmetric behavior of the QKPZ equation 
with respect to the sign of the nonlinear term. 
The QKPZ equation with $\lambda < 0$ in 1+1 dimensions has also been 
studied. We have obtained that the surface 
forms the shape of mountain with 
flat inclination and the PD transition is of first order.  
Since the anisotropic KPZ equation with thermal noise has been applied 
to the flux line dynamics \cite{hwa}, the AQKPZ equation 
considered in this letter may also be relevant to the 
flux line depinning problem in disordered media. 
Further details will be published elsewhere. \\

This work was supported in part by the KOSEF through the SRC 
program of SNU-CTP, by NON DIRECTED RESEARCH FUND, Korea 
Research Foundation, and in part by the Ministry of Education, Korea. \\

\begin{figure}
\centerline{\epsfxsize=7.3cm \epsfbox{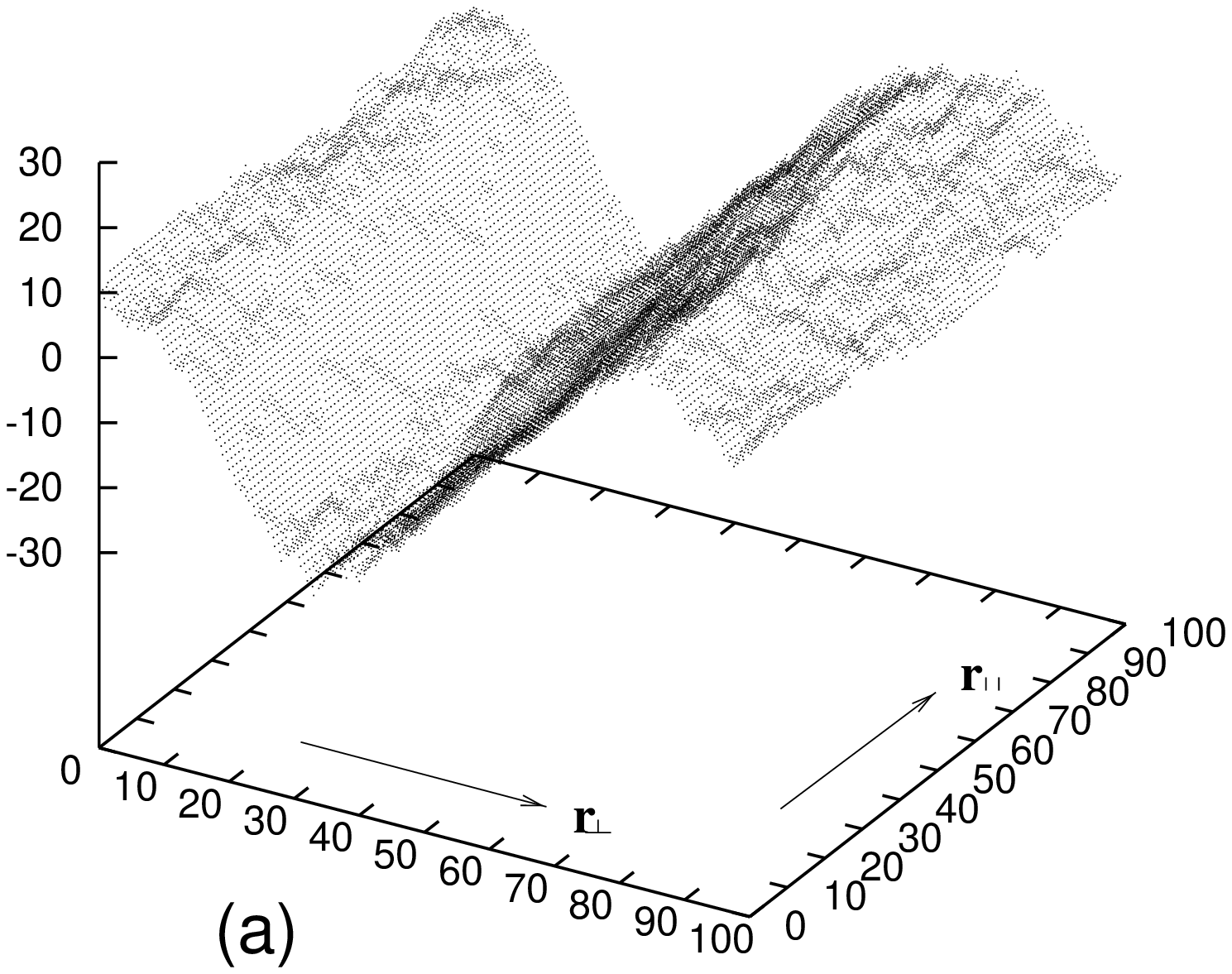}}
\centerline{\epsfxsize=7.3cm \epsfbox{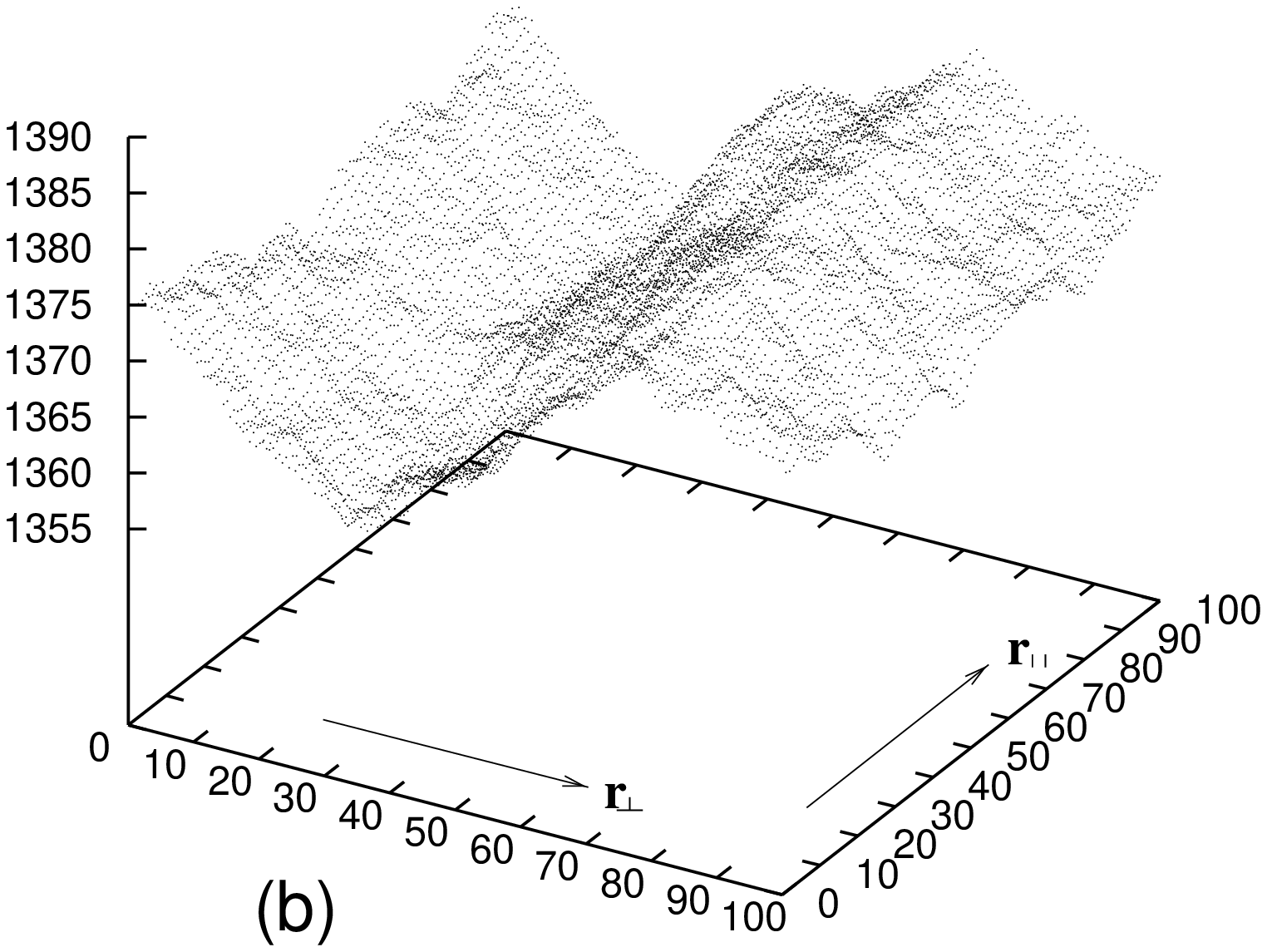}}
\caption{
Typical surface configuration of 
the AQKPZ equation generated by (a) the stochastic model and 
(b) direct numerical integration at the transition point.}
\label{fig1}
\end{figure}

\begin{figure}
\centerline{\epsfxsize=7.5cm \epsfbox{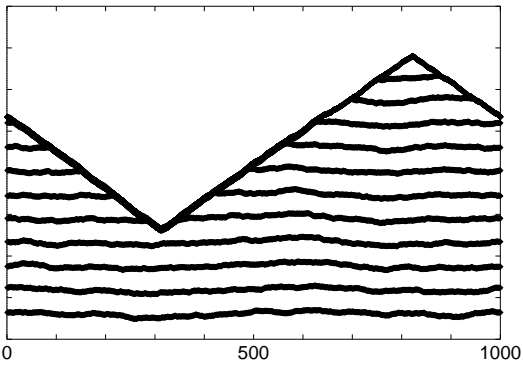}}
\vspace{.3cm}
\caption{
Temporally evolved surface configurations of the QKPZ 
equation with $\lambda <0$ in 1+1 dimensions at the 
pinning-depinning transition point. Successive height profiles are shown
at constant time intervals.}
\label{fig2}
\end{figure}

\begin{figure}
\centerline{\epsfxsize=7.5cm \epsfbox{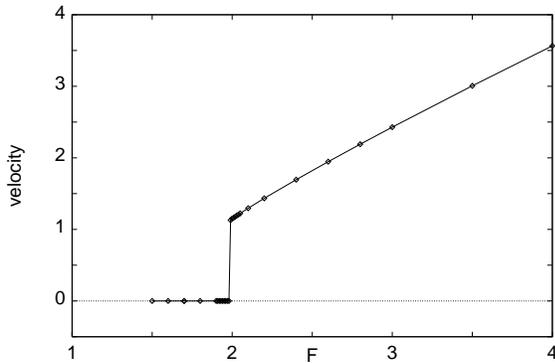}}
%\vspace{.2cm}
\caption{
The growth velocity versus force for the QKPZ equation 
with negative $\lambda$ in 1+1 dimensions.}
\label{fig3}
\end{figure}

\begin{figure}
\centerline{\epsfxsize=6cm \epsfbox{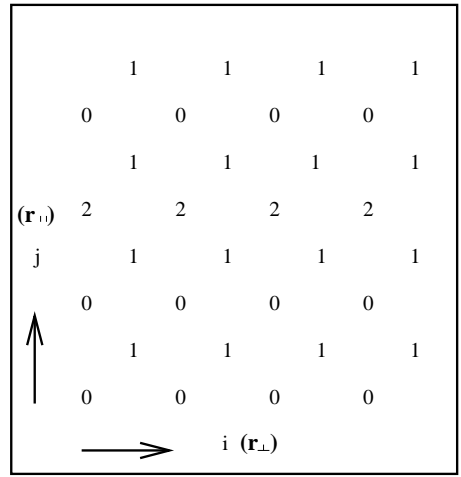}}
\vspace{.3cm}
\caption{
The configuration of a substrate with one step along 
a row for linear size $L=4$.
The avalanche process occurs in $\hat{j}$-direction.}
\label{fig4}
\end{figure}

\begin{figure}
\centerline{\epsfxsize=7.5cm \epsfbox{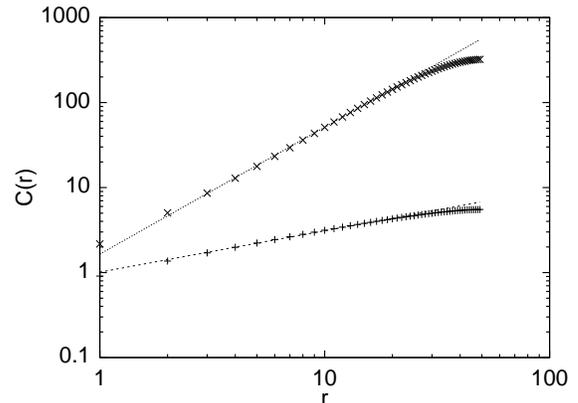}}
%\vspace{.2cm}
\caption{
The height-height correlation functions in parallel (lower data) and
perpendicular (upper ones) directions for the AQKPZ stochastic model
in 2+1 dimensions.}
\label{fig5}
\end{figure}

\begin{figure}
\centerline{\epsfxsize=7.5cm \epsfbox{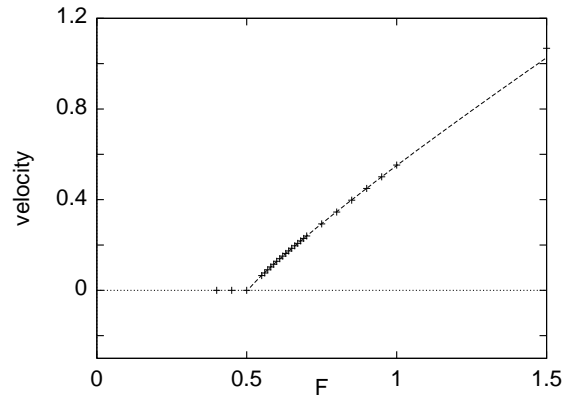}}
%\vspace{.2cm}
\caption{
The growth velocity versus force 
for the AQKPZ equation in 2+1 dimensions.}
\label{fig6}
\end{figure}

\end{multicols}
\end{document}